\documentclass[conference]{IEEEtran}

\usepackage{graphicx}
\usepackage{epstopdf}
\usepackage{standalone}
\usepackage[nolist ]{acronym}
\usepackage{tcolorbox}

\usepackage{pdfpages}
\usepackage{tikz}
\usetikzlibrary{positioning}
\usepackage{hyperref}
\usepackage{pdfcomment}
\usepackage{hologo}

\usepackage{xstring}

\usepackage[utf8]{inputenc}
\usepackage{marvosym}

\usepackage{algorithm}
\usepackage{algpseudocode}
\usepackage{tikz}

\usepackage{listings}
\definecolor{ListingBackground}{rgb}{0.97,0.97,0.97}
\usepackage{pgfplots}
\pgfplotsset{compat=newest}
\pgfplotsset{
    box plot/.style={
        /pgfplots/.cd,
        fill=blue!30,
        only marks,
        mark=-,
        mark size=0.2em,
        /pgfplots/error bars/.cd,
        y dir=plus,
        y explicit,
    },
    box plot box/.style={
        /pgfplots/error bars/draw error bar/.code 2 args={%
            \draw  ##1 -- ++(.2em,0pt) |- ##2 -- ++(-.2em,0pt) |- ##1 -- cycle;
        },
        /pgfplots/table/.cd,
        y index=2,
        y error expr={\thisrowno{3}-\thisrowno{2}},
        /pgfplots/box plot
    },
    box plot top whisker/.style={
        /pgfplots/error bars/draw error bar/.code 2 args={%
            \pgfkeysgetvalue{/pgfplots/error bars/error mark}%
            {\pgfplotserrorbarsmark}%
            \pgfkeysgetvalue{/pgfplots/error bars/error mark options}%
            {\pgfplotserrorbarsmarkopts}%
            \path ##1 -- ##2;
        },
        /pgfplots/table/.cd,
        y index=4,
        y error expr={\thisrowno{2}-\thisrowno{4}},
        /pgfplots/box plot
    },
    box plot bottom whisker/.style={
        /pgfplots/error bars/draw error bar/.code 2 args={%
            \pgfkeysgetvalue{/pgfplots/error bars/error mark}%
            {\pgfplotserrorbarsmark}%
            \pgfkeysgetvalue{/pgfplots/error bars/error mark options}%
            {\pgfplotserrorbarsmarkopts}%
            \path ##1 -- ##2;
        },
        /pgfplots/table/.cd,
        y index=5,
        y error expr={\thisrowno{3}-\thisrowno{5}},
        /pgfplots/box plot
    },
    box plot median/.style={
        /pgfplots/box plot
    },
    boxplot/every median/.style={
    	ultra thick,dashed,cyan
    }
}

\definecolor{flexicolor}{RGB}{46,49,146}
\definecolor{amaricolor}{RGB}{237,28,36}

\usepackage{xspace}

\usepackage[binary-units=true]{siunitx}
\usepackage{psfrag}
\usepackage{graphicx}
\usepackage{tabularx,booktabs}
\usepackage{multirow}
\usepackage{rotating}

\usepackage{multirow}

\usepackage[cmex10]{amsmath}
\usepackage[caption=false,font=footnotesize]{subfig}
\usepackage{stfloats}
\begin{document}

\newcommand{\paperTitle}{Towards Data-driven Simulation of End-to-end Network Performance Indicators}
\newcommand{\paperAuthors}{Benjamin Sliwa and Christian Wietfeld}
\newcommand{\paperEmails}{$\{$Benjamin.Sliwa, Christian.Wietfeld$\}$@tu-dortmund.de}

\newcommand{\figurePadding}{0pt}
\newcommand{\figureTopPadding}{\figurePadding}
\newcommand{\figureBottomPadding}{\figurePadding}

\newcommand{\dummy}[3]
{
	\begin{figure}[b!]  
		\begin{tikzpicture}
		\node[draw,minimum height=6cm,minimum width=\columnwidth]{\LARGE #1};
		\end{tikzpicture}
		\caption{#2}
		\label{#3}
	\end{figure}
}

\newcommand{\wDummy}[3]
{
	\begin{figure*}[b!]  
		\begin{tikzpicture}
		\node[draw,minimum height=6cm,minimum width=\textwidth]{\LARGE #1};
		\end{tikzpicture}
		\caption{#2}
		\label{#3}
	\end{figure*}
}

\newcommand{\basicFig}[7]
{
	\begin{figure}[#1]  	
		\vspace{#6}
		\centering		  
		\includegraphics[width=#7\columnwidth]{#2}
		\caption{#3}
		\label{#4}
		\vspace{#5}	
	\end{figure}
}
\newcommand{\fig}[4]{\basicFig{#1}{#2}{#3}{#4}{0cm}{0cm}{1}}

\newcommand{\subfig}[3]
{
	\subfloat[#3]{\includegraphics[width=#2\textwidth]{#1}}\hfill
}

\newcommand\circled[1] 
{
	\tikz[baseline=(char.base)]
	{
		\node[shape=circle,draw,inner sep=1pt] (char) {#1};
	}\xspace
}
\begin{acronym}
	\acro{KPI}{Key Performance Indicator}

	\acro{WEKA}{Waikato Environment for Knowledge Analysis}
	\acro{GPR}{Gaussian Process Regression}
	\acro{RF}{Random Forest}
	\acro{ANN}{Artificial Neural Network}
	\acro{SVM}{Support Vector Machine}
	\acro{M5}{M5 Regression Tree}
	\acro{CART}{Classification And Regression Tree}
	\acro{MDI}{Mean Decrease Impurity}
	
	\acro{RSRP}{Reference Signal Received Power} 
	\acro{RSRQ}{Reference Signal Received Quality} 
	\acro{SINR}{Signal-to-interference-plus-noise Ratio} 
	\acro{CQI}{Channel Quality Indicator} 
	\acro{ASU}{Arbitrary Strength Unit}
	\acro{TA}{Timing Advance}
	
	\acro{CAT}{Channel-aware Transmission}
	\acro{ML-CAT}{Machine Learning CAT}
	
	\acro{OMNeT++}{Objective Modular Network Testbed in C++}

	\acro{LTE}{Long Term Evolution}
	\acro{UE}{User Equipment}
	\acro{eNB}{evolved Node B}
	\acro{TCP}{Transmission Control Protocol}
	\acro{MUS}{Method Under Study}
	\acro{DDS}{Data-driven Simulation}
	\acro{TPC}{Transmission Power Control}
\end{acronym}

\acresetall
\title{\paperTitle}

\author{\IEEEauthorblockN{\textbf{\paperAuthors}}
	\IEEEauthorblockA{Communication Networks Institute,	TU Dortmund University, 44227 Dortmund, Germany\\
		e-mail: \paperEmails}}

\maketitle

%
%
\def\COPYRIGHTYEAR{2019}
\def\CONFERENCE{2019 IEEE 90th IEEE Vehicular Technology Conference (VTC-Fall)} 
\def\DOI{ 10.1109/VTCFall.2019.8891513}	

\def\bibtex
{
	@InProceedings\{Sliwa/Wietfeld/2019b,\\
	author    = \{Benjamin Sliwa and Christian Wietfeld\},\\
	title     = \{Towards data-driven simulation of end-to-end network performance indicators\},\\
	booktitle = \{2019 IEEE 90th Vehicular Technology Conference (VTC-Fall)\},\\
	year      = \{2019\},\\
	address   = \{Honolulu, Hawaii, USA\},\\
	month     = \{Sep\},\\
	\}
}
\ifx\CONFERENCE\VOID
\def\conferencenotice{Submitted for publication}
\def\copyrightnotice{}
\else
\ifx\DOI\VOID
\def\conferencenotice{Accepted for presentation in: \CONFERENCE}	
\else
\def\conferencenotice{Published in: \CONFERENCE\\DOI: \href{http://dx.doi.org/\DOI}{\DOI}

}
\fi
\def\copyrightnotice{
	\copyright~\COPYRIGHTYEAR~IEEE. Personal use of this material is permitted. Permission from IEEE must be obtained for all other uses, including reprinting/republishing this material for advertising or promotional purposes, collecting new collected works for resale or redistribution to servers or lists, or reuse of any copyrighted component of this work in other works.
}
\fi
\def\overlayimage{%
	\begin{tikzpicture}[remember picture, overlay]
	\node[below=5mm of current page.north, text width=20cm,font=\sffamily\footnotesize,align=center] {\conferencenotice \vspace{0.3cm} \pdfcomment[color=yellow,icon=Note]{\bibtex}};
	\node[above=5mm of current page.south, text width=15cm,font=\sffamily\footnotesize] {\copyrightnotice};
	\end{tikzpicture}%
}
\overlayimage
\begin{abstract}
	
%
%
Novel vehicular communication methods are mostly analyzed simulatively or analytically as real world performance tests are highly time-consuming and cost-intense. Moreover, the high number of uncontrollable effects makes it practically impossible to reevaluate different approaches under the exact same conditions.
%
%
However, as these methods massively simplify the effects of the radio environment and various cross-layer interdependencies, the results of end-to-end indicators (e.g., the resulting data rate) often differ significantly from real world measurements.
%
%
In this paper, we present a data-driven approach that exploits a combination of multiple machine learning methods for modeling the end-to-end behavior of network performance indicators within vehicular networks. The proposed approach can be exploited for fast and close to reality evaluation and optimization of new methods in a controllable environment as it implicitly considers cross-layer dependencies between measurable features.
%
%
Within an example case study for opportunistic vehicular data transfer, the proposed approach is validated against real world measurements and a classical system-level network simulation setup. 
%
%
Although the proposed method does only require a fraction of the computation time of the latter, it achieves a significantly better match with the real world evaluations.

\end{abstract}

\IEEEpeerreviewmaketitle

\section{Introduction}

%
%
Designing and evaluating novel communication mechanisms for vehicular networks is a highly complex task as the performance of the radio link is severely impacted by mobility-related effects, the properties of the environment and interference \cite{Sliwa/etal/2019b}. Since comprehensive real world evaluations are very time-consuming, it is often not feasible to perform parameter studies in the real world.

%
%
Therefore, system-level network simulation is a commonly used approach, which enables to analyze various parameterizations and to evaluate novel schemes in different scenarios. In contrast to field tests, one of the main strengths of simulations is their system-immanent control of the environment, which allows to evaluate different methods under the exact same conditions. However, as the real world radio medium is shared with other cell users, for which the communication pattern is unknown, the generation of realistic simulation setups is itself a highly challenging -- if not impossible -- task \cite{Cavalcanti/etal/2018a}. Therefore, the results of these simulations differ significantly from real world evaluations.

%
%
The analysis of end-to-end network performance indicators (e.g., the resulting data rate) within mobile networks is even more complicated, as it is impacted by various cross-layer dependencies within the device itself, mobility-related phenomenons and effects related to the coexistence of different users and technologies in a shared radio medium (see Fig.~\ref{fig:scenario}) \cite{Sliwa/Wietfeld/2019b}.
%
%
\basicFig{!b}{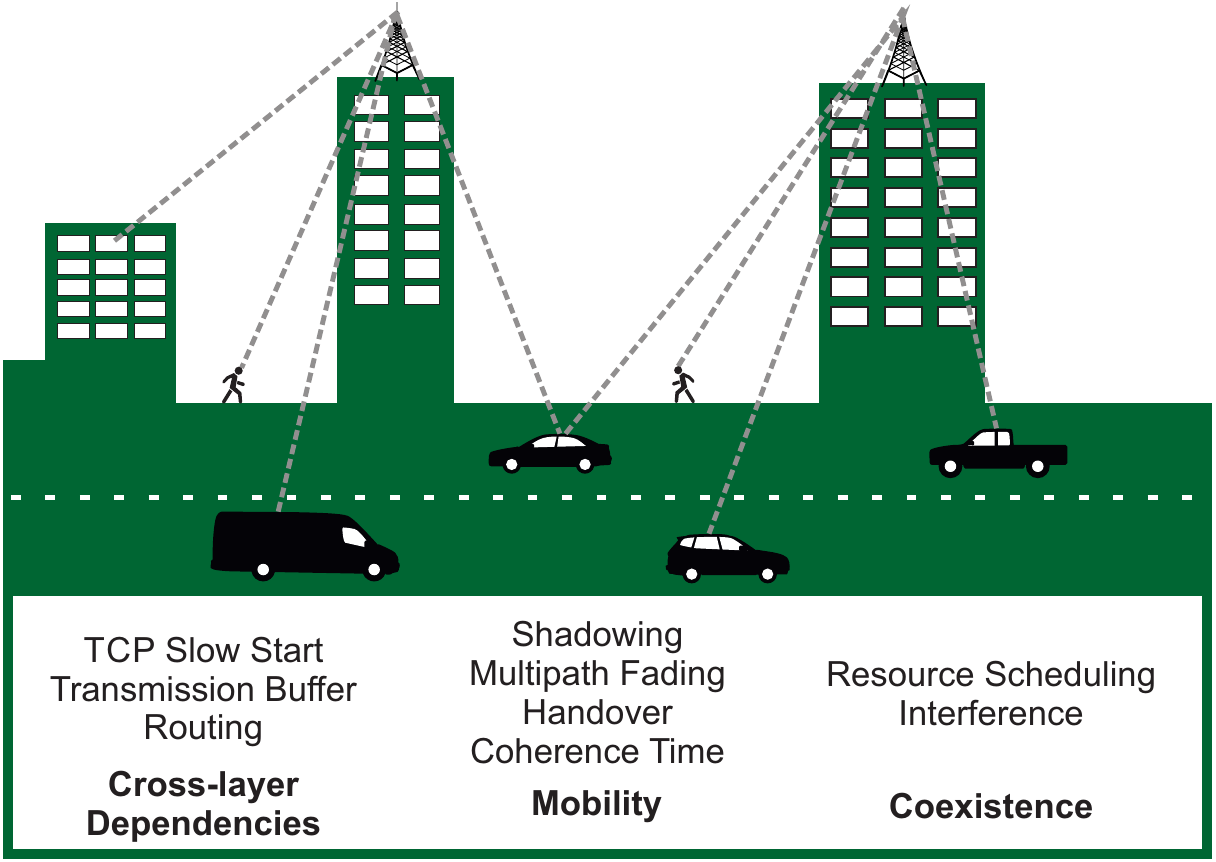}{Overview about different effects, which have an impact on the resulting end-to-end behavior of cellar vehicular networks. The strength of machine learning is the implicit consideration of hidden interdependencies, which are partly covered by measurable indicators.}{fig:scenario}{0cm}{0cm}{0.9}
%
%
In this paper, we present a fundamentally different alternative to established analysis methods, which relies on a data-driven approach for machine learning-based modeling of end-to-end network performance indicators. In contrast to system-level network simulation, which is mainly intended for the \emph{system conception phase}, the proposed approach targets the \emph{system optimization phase} (see Fig.~\ref{fig:dds}). Real world measurement data is exploited in an automated process to derive a \emph{prediction model}, which is suitable for simulative evaluation as well as for online application within live systems. In addition, a second machine learning model is derived based on \ac{GPR}, which learns the statistical \emph{distribution of the prediction errors}. Both models are brought together in a lightweight simulation setup -- which is referred to as \ac{DDS} -- for analyzing a novel \ac{MUS}.

%
%
With the proposed approach, novel communication methods can be analyzed under close to reality conditions in a fast and resource-efficient way. Overall, the proposed machine learning-based method brings together the accuracy of real world experiments, the environment control of simulation setups and the evaluation speed of analytical modeling.
%
%
The contributions provided by this paper are as follows:
\begin{itemize}
	\item An automated process for generating prediction and error models of end-to-end network performance indicators based on a \textbf{combination of multiple machine learning methods}.
	\item Proof-of-concept evaluation and \textbf{validation against system-level simulation and real world measurements} in a case study for anticipatory transmission of vehicular sensor data. 
	\item The framework itself and the raw results of the considered case study are provided using an \textbf{Open Access} approach.
\end{itemize}
%
%
The remainder of the paper is structured as follows. After discussing relevant state-of-the-art approaches in Sec.~\ref{sec:related_work}, we present the system model of the proposed solution approach in Sec.~\ref{sec:approach}. Afterwards, a case study for opportunistic data transmission is discussed, where the proposed \ac{DDS} is validated based on real world measurements and compared to a purely simulative approach in Sec.~\ref{sec:results}. Finally, the limitations of \ac{DDS} are discussed in Sec.~\ref{sec:limitations}.
%
%
\fig{b}{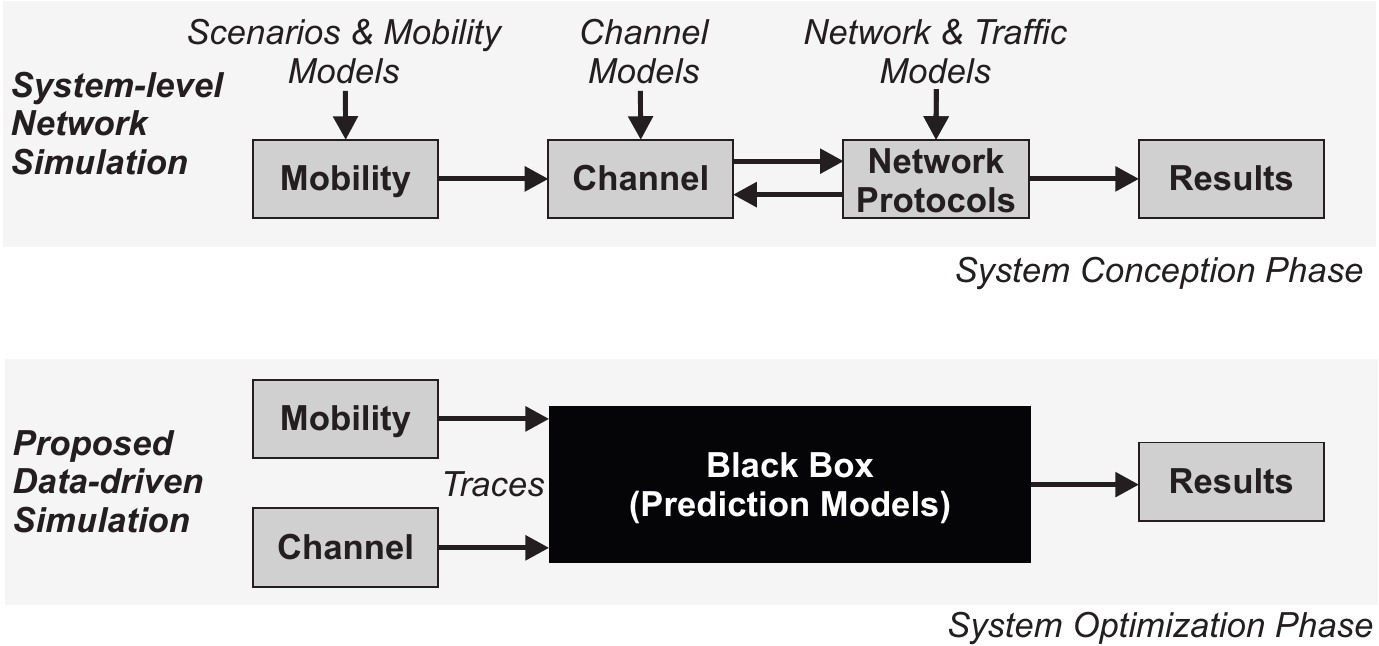}{Comparison of the proposed \ac{DDS} approach to classical system-level network simulation.}{fig:dds}
Fig.~\ref{fig:dds}
\section{Related Work} \label{sec:related_work}

%
%
In recent years, machine learning has become an essential method for analyzing and optimizing communication systems and processes. Comprehensive surveys about methods and applications of machine learning within mobile wireless networks are provided by \cite{Ye/etal/2018a} and \cite{Jiang/etal/2017a}.

%
%
In \cite{Cavalcanti/etal/2018a}, the authors present the results of a comprehensive literature review campaign with the aim of determining the most important topics, methods and trends of vehicular communication networks. One of the main findings is that many of the considered research works rely on unrealistic simulation models for both communication and mobility.
%
%
Although the accuracy of simulation results can be increased by using more detailed -- and more computation-intense -- simulation models \cite{Mir/2019a}, the required high-resolution data is often not available (e.g., material parameters for ray tracing simulation \cite{Yun/Iskander/2015a}). As a result, system-level network simulation is rather suitable for comparing novel methods in abstract environments than to derive conclusions for defined real world scenarios \cite{Cavalcanti/etal/2018a}.

%
%
Anticipatory communication \cite{Bui/etal/2017a} is a novel wireless networking paradigm, which aims to optimize the performance of communication systems by integrating context information into decision processes. It is especially well-suited for systems that experience highly dynamic radio conditions (e.g., due to the mobility behavior within vehicular networks).
As anticipatory communication often aims to optimize a defined number of end-to-end performance indicators, it requires methods to describe and forecast the latter.
%
%
In previous work, we have demonstrated the massive potentials of this approach for improving reliable mesh routing by mobility prediction \cite{Sliwa/etal/2016a} and optimizing the resource efficiency of vehicular sensor data transfer \cite{Sliwa/etal/2018a, Sliwa/etal/2018b, Sliwa/etal/2019d} by machine learning-based data rate prediction.
%
%
One of the key strengths of these machine learning-enabled approaches is their \emph{data driven} description of the environment. Instead of actually modeling a scenario with different interacting entities, only the interdependencies of multiple measurable variables are considered.

%
%
Within this work, we exploit the insights in data-driven modeling obtained by anticipatory communication to compensate the shortcomings of classical simulative evaluation. Recently, first results for learning end-to-end models of communication systems have been proposed by Ye et al. and \cite{Ye/etal/2018b} and Aoudia et al. \cite{Aoudia/Hoydis/2018a}. While these works analyze the physical layer modeling, the proposed approach focuses on the application layer behavior.

\section{Machine Learning-based Solution Approach} \label{sec:approach}

%
%
\basicFig{}{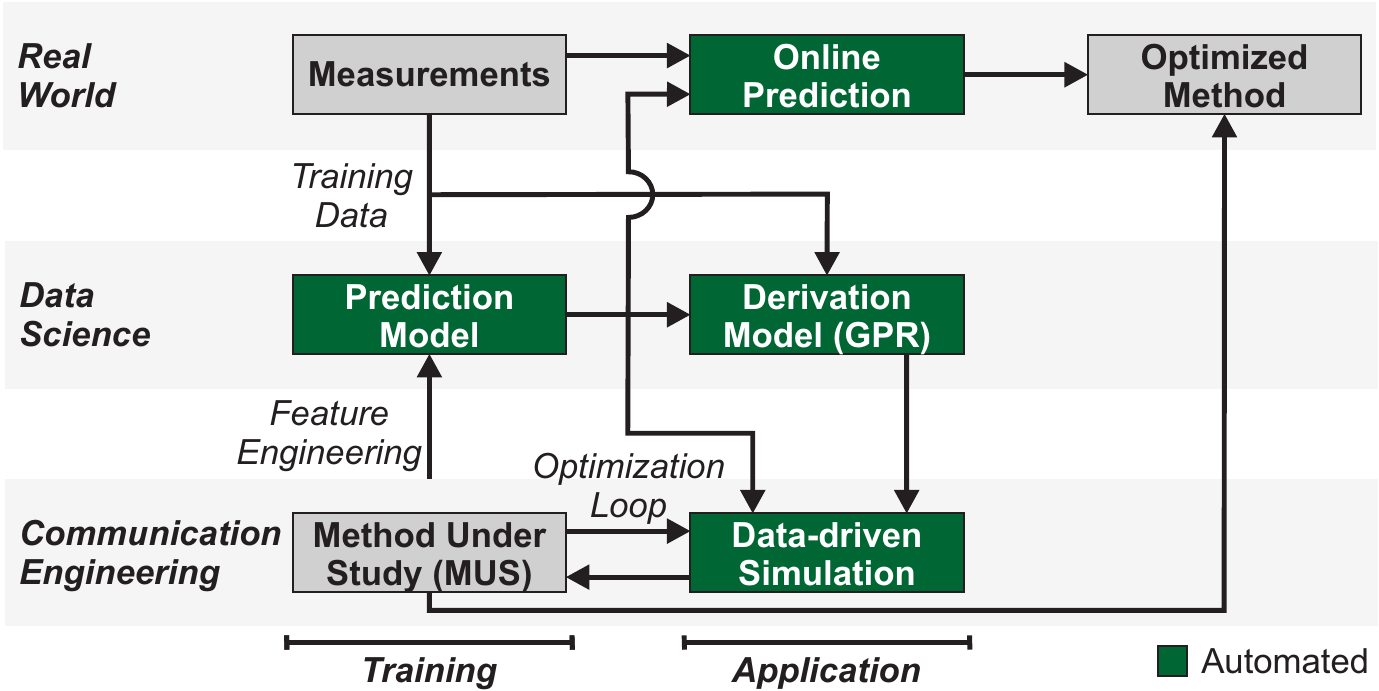}{System architecture model of the proposed multi-disciplinary machine learning-based analysis framework.}{fig:approach}{-0.5cm}{0cm}{1}
The overall architecture of the proposed solution approach is shown in Fig.~\ref{fig:approach}. The framework operates on three different logical domains: \emph{real world}, \emph{data science} and \emph{communication engineering}. The overall goal is to optimize a novel \ac{MUS} using \ac{DDS} and bring it from theory to real world application.
%
%
Although many of the components are derived by means of automation, the framework relies on two manual steps as prerequisites. In order to enable the data-driven modeling approach, a comprehensive data set needs to be obtained in the real world, which should cover all external influences on the considered \ac{KPI} as close as possible. Furthermore, expert knowledge from the communication engineering domain is required for the feature engineering phase, in order to monitor meaningful variables that describe the behavior of the considered end-to-end indicator.
%
%
After the real world data acquisition, a \ac{DDS} model is composed by the \ac{MUS}, a derived prediction model (see Sec.~ \ref{sec:prediction_model}) and a derived machine learning-based model for the derivation behavior of the prediction model itself (see Sec.~ \ref{sec:derivation_model}). Within an optimization loop, the \ac{MUS} can now be analyzed and optimized based on \ac{DDS}.
 
%
%
Finally, the optimized \ac{MUS} can be transfered to a real world implementation. As the generated prediction model is represented as raw \texttt{C++} code, the trained model can be directly exploited for performing online predictions in the real world deployment.

\subsection{Generating and Analyzing Machine Learning-based Prediction Models} \label{sec:prediction_model}

%
%
Network quality prediction is a regression task, which uses supervised learning based on a labeled data set $\mathbf{Y}$ (measurements of the end-to-end network performance indicator) and a training set $\mathbf{X}$ (measurements of related indicators). After the training phase, the trained model $f$ can be utilized to make predictions $\tilde{y}$ for unlabeled features sets $\mathbf{x}$ such that $\tilde{y} = f(\mathbf{x})$. Within the proposed framework, this step is performed in an automated way, which is further described in Sec.~\ref{sec:methods}.

%
%
As different evaluations have shown (e.g., \cite{Sliwa/etal/2018b, Samba/etal/2017a, Jomrich/etal/2018a}), for predicting network quality indicators, the application of \ac{CART}-based models -- e.g., \ac{RF} \cite{Breiman/2001a} and \ac{M5} \cite{Quinlan/1992a} -- often yields better performance than more complex regression models like \ac{ANN} and \ac{SVM}. 
In many cases, one of the considered network quality indicators has a dominant impact on the behavior of the considered \ac{KPI} within a defined scope.
As an example for \ac{LTE}, the analysis in \cite{Sliwa/etal/2018b} points out that at the cell edge -- which can be identified by the \ac{RSRP} -- the interference level, which is partly identifiable by the \ac{RSRQ}, has a strong impact on the data rate. In contrast to that, the \ac{SINR} is of higher importance within the cell center. This segment-wise definition of ranges of validity matches the general structure of the \ac{CART} models.
%
%
In addition to their high prediction accuracy, one of the main advantages of these models is that their tree-like structure allows a resource-efficient implementation using a sequence of \texttt{if/else} statements.

%
%
It needs to be remarked that the accuracy of the resulting model is severely depending on the quality and the amount of the training data. Therefore, feature engineering is of tremendous importance in order to achieve a reusable model that works accurately within the considered evaluation scenario and provides a good generalization.
 
%
%
However, in contrast to system-level simulation, which requires complex evaluation setup phase, the derived model is suitable for making predictions in various scenarios after a sufficient amount of data has been obtained. 
%
%
For \ac{CART}-based prediction models, the tree structure can be exploited to visually assess the impact of individual features. In addition, methods such as \ac{MDI} \cite{Louppe/etal/2013a} allow to get a deeper insight into the importance of individual features within \acp{RF}.

%
%
For comparing the resulting prediction quality of different models, the \emph{coefficient of determination} $R^{2}$ provides a statistical metric that describes how accurate measurement values fit to a regression model. It is calculated as 
%
%
\begin{eqnarray}
	R^{2} = 1- \frac{\sum_{i=1}^{N}\left(\tilde{y}_{i} - y_{i} \right)^{2}}{\sum_{i=1}^{N}\left(\bar{y} - y_{i} \right)^{2}}
\end{eqnarray}
with $\bar{y}_{i}$ being the mean measurement value, $y_{i}$ being the current measurement and $\tilde{y}_{i}$ being the current prediction.

\subsection{Machine Learning-based Modeling of Derivations Between Prediction Model and Real World Measurements} \label{sec:derivation_model}

The derived regression model can now be utilized to make predictions for unobserved values. However, for each input feature set $\mathbf{x}$, only a single predicted value $\tilde{y}$ is obtained, which is constant if the same feature set is evaluated again. 

%
%
Derivations within the distribution of the measured values are related to other real world influences, which are not considered by the model. However, in the targeted data-driven simulation, the results should represent the real world behavior and not just its predicted average. 
Therefore, a second machine learning model is applied for learning the characteristics of the derivations -- which are assumed to have a gaussian distribution -- between predictions and real world measurements. For this task, a \ac{GPR} \cite{Rasmussen/2004a} is \emph{trained on the results of the prediction model} such that $\tilde{\mathbf{Y}}_\text{GPR} = f_{\text{GPR}}(\tilde{\mathbf{Y}})$.
A gaussian process $\mathcal{GP}(m,k)$ is a collection of random variables with a joint gaussian distribution and is described by its mean function $m$ and its covariance function $k$. This method is considered \emph{non-parametric}, as it learns a description for the data as a distribution over possible functions (in contrast to regular regression, which learns the distribution over function parameters).
%
%
\begin{figure}[]
	\centering		  
	\includegraphics[width=1\columnwidth]{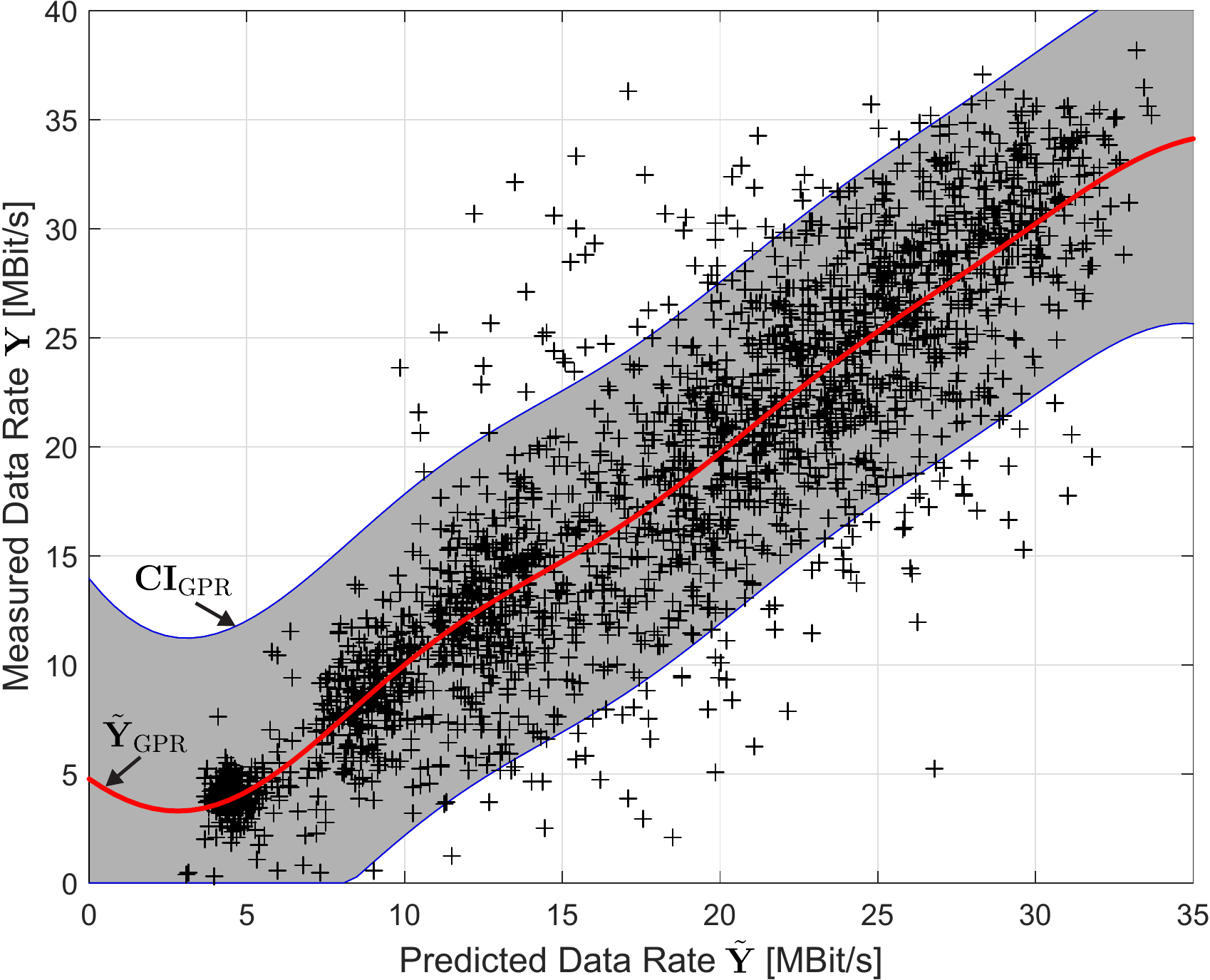}
	\vspace{-0.5cm}	
	\caption{For consideration of the error distribution of the trained prediction model, a \ac{GPR} is trained on the prediction results of the prediction model itself. The shaded area illustrates the posterior confidence region.}	
	\label{fig:prediction_gpr}
	\vspace{-0.5cm}	
\end{figure}
As illustrated by the example in Fig.~\ref{fig:prediction_gpr}, the \ac{GPR} does not only compute predictions $\tilde{\mathbf{Y}}_\text{GPR}$ but also derives a confidence interval $\mathbf{CI}_\text{GPR}$ and a standard deviation function $\boldsymbol{\sigma}_\text{GPR}$. Therefore, we can now compute a randomized sample $\tilde{y}_{\text{GPR}}$, which follows a gaussian distribution $\mathcal{N}$ of the prediction errors as
%
%
\begin{equation}
	\tilde{y}_{\text{GPR}} 
	= \mathcal{N}\left( 
	\tilde{\mathbf{Y}}_\text{GPR}(\tilde{y}), 
	\boldsymbol{\sigma}_\text{GPR}^2(\tilde{y})
	\right)
\end{equation}

%
%
For the data shown in Fig.~\ref{fig:prediction_gpr}, it can be seen that the confidence area for $\tilde{\mathbf{Y}}<3~\text{MBit/s}$ and $\tilde{\mathbf{Y}}>33.5~\text{MBit/s}$ is significantly larger than for the values in between. In these edge regions, the prediction model is not sufficiently covered by the supplied training data and therefore the resulting prediction accuracy is low.
As a consequence to these statistical effects, unrealistic or even impossible values (e.g., negative data rates) can occur for $\tilde{y}_{\text{GPR}}$ in these regions. To avoid this behavior, the results are \emph{shaped to the value range of the measurements} in a final step with
%
%
\begin{equation}
	\hat{y} = \begin{cases}
	\min(\mathbf{Y}) & \tilde{y}_{\text{GPR}} < \min(\mathbf{Y}) \\
	\max(\mathbf{Y}) & \tilde{y}_{\text{GPR}} > \max(\mathbf{Y}) \\
	\tilde{y}_{\text{GPR}}  & \text{else}
	\end{cases}
\end{equation}
%
%
For completeness, it is remarked that different machine learning models (e.g., the considered \ac{GPR}) are able to derive confidence regions based on the input features themselves. However, the advantage of the presented separation of prediction model and derivation model is its modularity. As the \ac{GPR} only relies on the measurements $\mathbf{Y}$ and the prediction results $\tilde{\mathbf{Y}}$, the prediction model itself can be easily replaced by any other regression model.

\section{Validation} \label{sec:results}

In this section, we present a proof-of-concept evaluation of the proposed \ac{DDS} approach and validate it against reference system analysis methods. 
%
%
As \acp{MUS}, we study the opportunistic sensor data transmission schemes \ac{CAT} and the \ac{ML-CAT} \cite{Sliwa/etal/2018a, Sliwa/etal/2018b}, which only have been analyzed empirically in previous work. 
 
%
%
For the validation, the algorithms of the considered \acp{MUS} are implemented in three methodological setups: the proposed \ac{DDS}, a classical system-level network simulator and in a real world system. The goal is to analyze how well the proposed \ac{DDS} and the considered reference methods are able to represent the real world behavior of the transmission schemes.

\subsection{Method Under Study: \ac{CAT}-based Data Transfer}

Straightforward periodic transmission methods within cellular vehicular networks do not consider the network quality within the transmission process itself. Therefore, cell- and energy resources are often wasted as data transmissions during low channel quality periods suffer from packet loss and lead to retransmissions.

%
%
To overcome these issues, \ac{CAT} and \ac{ML-CAT} \cite{Sliwa/etal/2018a, Sliwa/etal/2018b} have been proposed, which rely on a probabilistic process to compute a transmission probability $p(t)$ with respect to the current network conditions. Data is stored in a local buffer until a transmission decision is made for the whole buffer.  
%
%
As this approach introduces an additional delay -- the transmission scheme waits for better channel conditions -- it is intended for delay-tolerant applications (e.g., intelligent traffic control, predictive maintenance, environmental monitoring) and not suitable for safety-critical real time applications.
The resulting transmission probability is computed as 
%
%
\begin{equation}
	p(t) = \begin{cases}
		0 & \Delta t < t_{\min} \\
		1 & \Delta t > t_{\max} \\
		\left( \frac{\Phi(t)-\Phi_{\min}}{\Phi_{\max}-\Phi_{\min}} \right)^\alpha  & \text{else} \\
	\end{cases}
\end{equation}
with the exponent $\alpha$ for controlling the preference of high metric values and $\Delta t$ being the passed time since the last transmission has been performed . $t_{\min}$ is used to guarantee a minimum packet size and $t_{\max}$ specifies an upper bound for the buffering delay, which has to be chosen with respect to the application requirements.

%
%
For the validation, an \ac{SINR}-based metric $\Phi_{\text{SINR}} = \left\lbrace 0~\text{dB}, 30~\text{dB} \right\rbrace$ and a throughput prediction-based metric $\Phi_{\text{RF}} = \left\lbrace 0~\text{MBit/s}, 30~\text{MBit/s} \right\rbrace$ are applied. For comparison, straightforward periodic data transmissions with a fixed interval of $\Delta t=10~s$ are considered.

\subsection{Methodological Setup} \label{sec:methods}

For the validation, the considered variants of \ac{CAT} are analyzed using \ac{DDS} and are compared to real world evaluations, which are carried out in different drive tests the public \ac{LTE} network. An Android-based \ac{UE} (Samsung Galaxy S5 Neo, Model SM-G903F) executes the \ac{CAT} process, which models a virtual sensor application. Data is transfered via \ac{TCP} in the uplink from the \ac{UE} through the cellular network to a cloud-based server, which measures the resulting data rate. 
%
%
Per configuration, five drive tests are performed on a 9~km long \emph{suburban} track as well as on a 14~km long \emph{highway} track.
Tab.~\ref{tab:parameters} summarizes further key parameters of the evaluation setup.
%
%
\newcommand{\entry}[2]{&#1 & #2 \\}
\newcommand{\head}[2]{& \toprule \entry{\textbf{#1}}{\textbf{#2}}}

\newcommand{\sideHeader}[3]
{
	\multirow{#1}{*}{
		\rotatebox[origin=c]{90}{
			\parbox{#2}{\centering \textbf{#3}}
		}
	}
}

\begin{table}[ht]
	\centering
	\caption{Validation Parameters}
	\begin{tabular}{p{0.2cm}p{3.5cm}p{3.8cm}}
		\toprule
		\sideHeader{11}{0.5cm}{General} 
		\entry{\textbf{Parameter}}{\textbf{Value}}
		\midrule
		\entry{Data source}{50~kByte/s}
		\entry{Evaluation interval}{1~Hz}
		\entry{$t_{\min}$}{10~s}
		\entry{$t_{\max}$}{120~s}
		\entry{$\alpha$}{6}
		\entry{Metric for \ac{CAT}}{$\Phi_{\text{SINR}} = \left\lbrace 0~\text{dB}, 30~\text{dB} \right\rbrace$}
		\entry{Metric for \ac{ML-CAT}}{$\Phi_{\text{RF}} = \left\lbrace 0~\text{MBit/s}, 30~\text{MBit/s} \right\rbrace$}

		\midrule
		
		\sideHeader{8}{0.3cm}{SimuLTE} 
		\entry{Carrier frequency}{2100~MHz}
		\entry{Bandwidth}{20~Mhz}
		\entry{\ac{UE} transmission power}{23~dBm}
		\entry{\acs{eNB} transmission power}{43~dBm}
		\entry{Channel model}{WINNER II Urban Macrocell}

		\bottomrule
		
	\end{tabular}
	\label{tab:parameters}
\end{table}

%
%
The machine learning based data rate prediction metric computes its current values as $\Phi_{\text{RF}}(t)=f_{\text{RF}}(\mathbf{x}(t))$ with $\mathbf{x}(t)$ being a set of ten measured features: payload size, \ac{RSRP}, \ac{RSRQ}, \ac{SINR}, \ac{CQI}, \ac{ASU}, \ac{TA}, carrier frequency, cell id, and velocity. The prediction model $f_{\text{RF}}$ is a \ac{RF}, which consists of 100 random trees and is trained on the open data set of \cite{Sliwa/2019a}, which contains the results and features of 2342 uplink transmissions.
%
%
On the supplied data set, the prediction model achieves a coefficient of determination of $R^2 = 0.8$ after 10-fold cross validation. For training the actual machine learning model, the \ac{WEKA} toolkit \cite{Hall/etal/2009a} is used. In order to automate the training phase and to generate implementations for performing online predictions, an interface application has been developed\footnote{The software is available at https://github.com/BenSliwa/DDS}.

%
%
As a methodological reference, a simulation setup in \ac{OMNeT++}~5.0 \cite{Varga/Hornig/2008a} with INET~3.4 and SimuLTE~v0.9.1 \cite{Virdis/etal/2015a} is analyzed, which replays the real world measurements. The \acp{eNB} are positioned with respect to their real world locations and the analyzed \ac{UE} moves according to the recorded trajectory. 
%
%
Within the simulation setup, the prediction model $f_{\text{RF}}$ cannot be directly implemented as the simulator only implements a fraction of the considered features. Therefore, a reduced prediction model $\tilde{y}^{*}_{RF} = f^{*}_{RF}(\mathbf{x^{*}})$ is derived, which utilizes the available features payload size and the \ac{SINR} for the online data rate prediction.
%
%
\basicFig{}{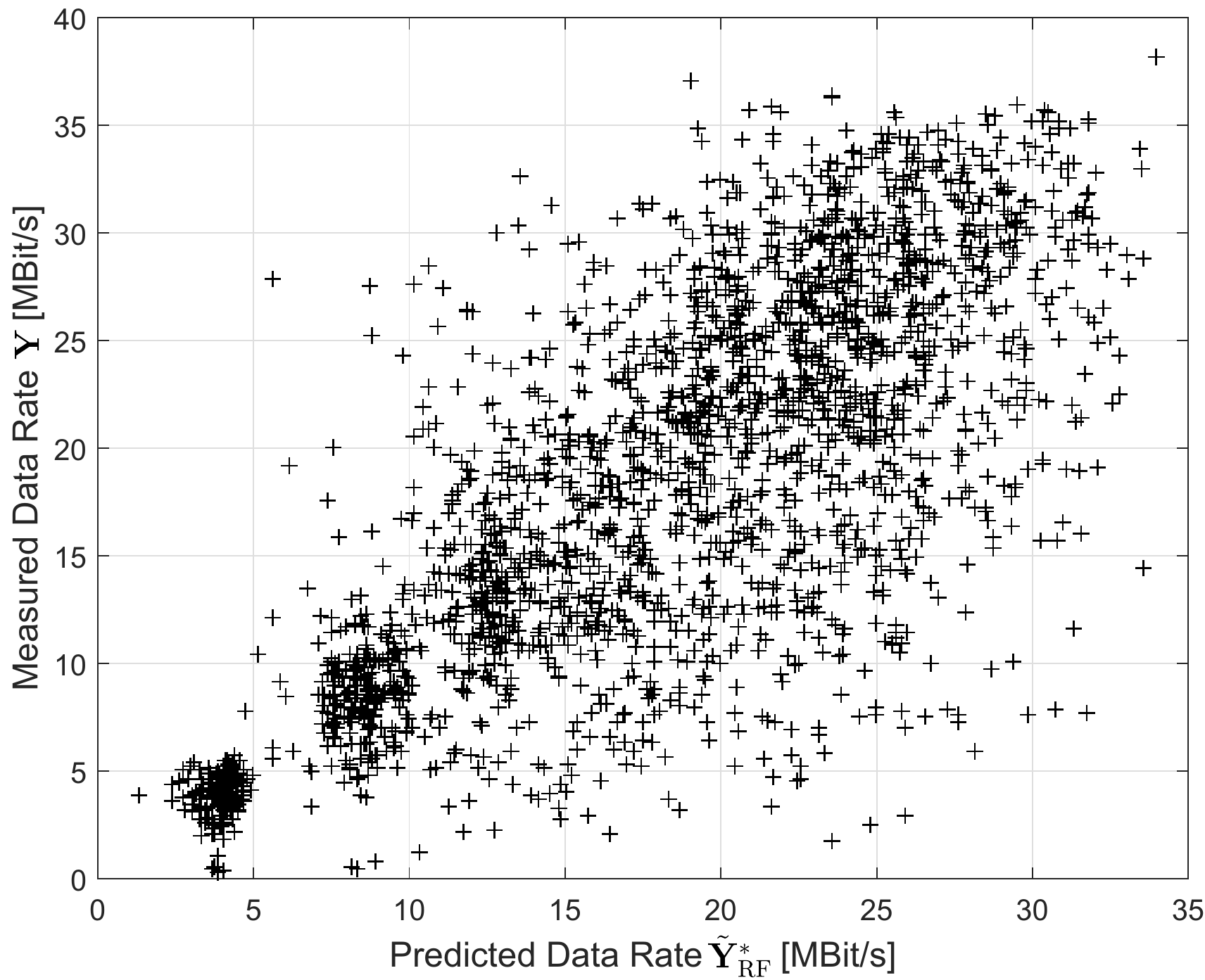}{Resulting data rate prediction performance for the simplified model $f^{*}_{\text{RF}}$, which is applied in the system-level network simulation.}{fig:rf_simplified}{-0.5cm}{0cm}{1}
As a consequence to the model simplification, the resulting prediction accuracy is significantly reduced ($R^2 = 0.5655$), which can be observed in Fig.~\ref{fig:rf_simplified}.
%
%
All measurements are executed with an Intel Core i5-6200@2.3 GHz, 8 GB RAM and Ubuntu 18.04.1 LTS operating system.

\subsection{Results}

In the following paragraph, the results of different individual evaluations are presented and discussed.

\subsubsection{Modeling the Distribution of Real World Measurements}

Before the results for the \ac{CAT} transmissions are analyzed, the impacts of the prediction and the \ac{GPR}-based derivation model (see Sec.~\ref{sec:approach}) are studied.
For this, all 2342 data transmissions of the training set are repeated by means of \ac{DDS} and compared to the achieved real world results. For each transmission, the recorded features are utilized by the \ac{RF}-based prediction.
%
%
\basicFig{}{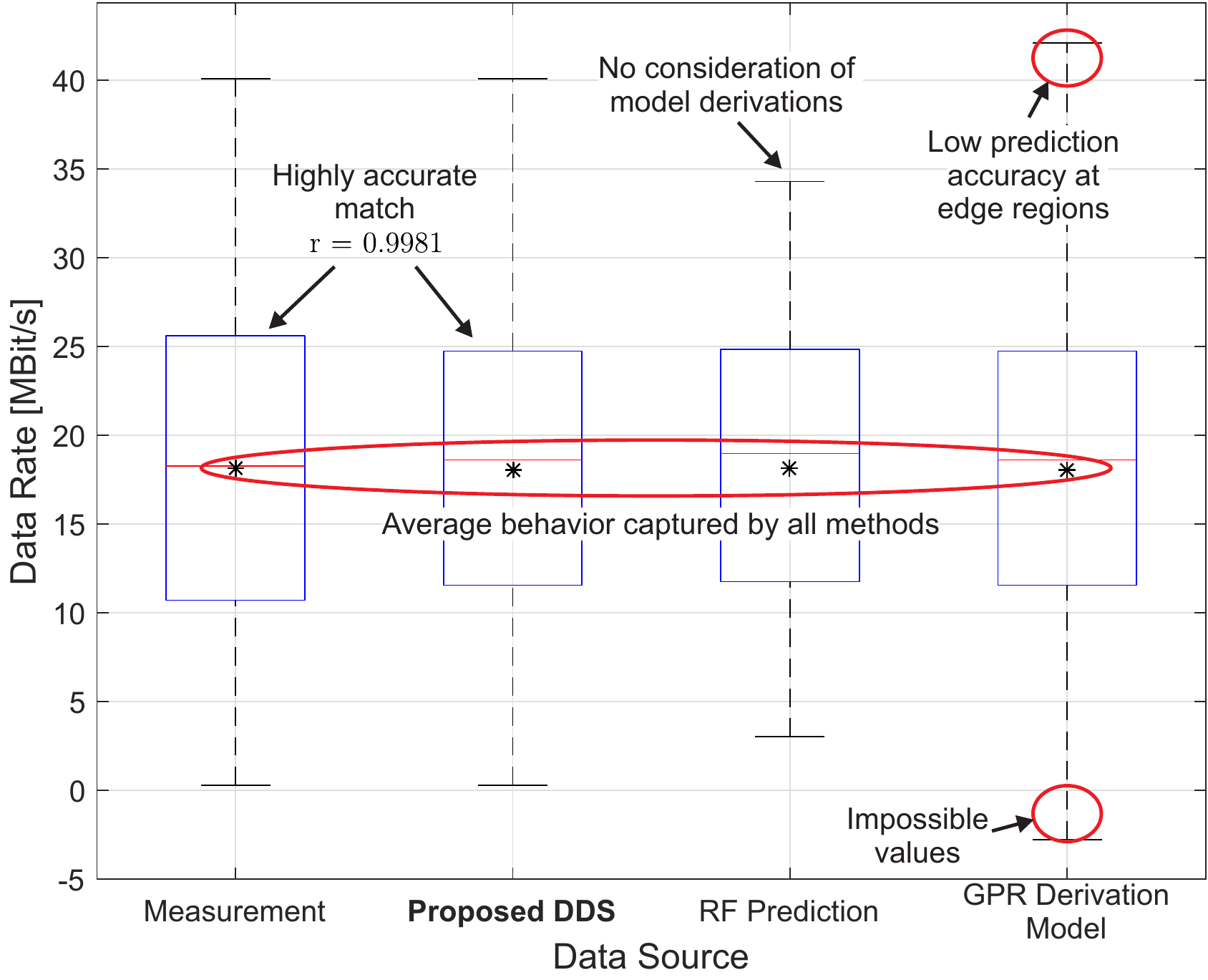}{Comparison of the data distribution modeling accuracy. The goal of each method is to achieve the highest congruency with the real world measurements.}{fig:gpr_box}{-0.5cm}{0cm}{1}
Fig.~\ref{fig:gpr_box} shows a comparison of the resulting accuracy for modeling the distribution of the measurement values. While all approaches are able to cover the average behavior of the observations, the resulting distributions show differences in the distribution of the data rate values.
%
%
The \ac{RF} prediction model does not consider model derivations, therefore the resulting value range is centered narrowly around the mean value of the measurements. 
%
%
In contrast to that, the raw \ac{GPR} approach overemphasizes the statistical uncertainties in the edge regions. As a consequence, even impossible values occur (negative data rates)

%
%
Finally, \ac{DDS} brings together both components and is able to detect unrealistic derivations from the observed reality. The results show a high match with the real world measurements with a correlation coefficient of $r=0.9981$.

\subsubsection{End-to-end Performance of the \acp{MUS}}

%
%
It is important to remark that \ac{DDS} and system-level simulation utilize the data of the \emph{training set} to replay the trajectory of the vehicle and in the case of \ac{DDS}, also the channel conditions. The real world validation itself is performed against \emph{additional measurements}, which are not part of the training set and are only used for validation. This data is captured on the same physical tracks but subject to different mobility and channel-related effects.
%
%
\begin{figure*}[] 
	\centering
	
	\subfig{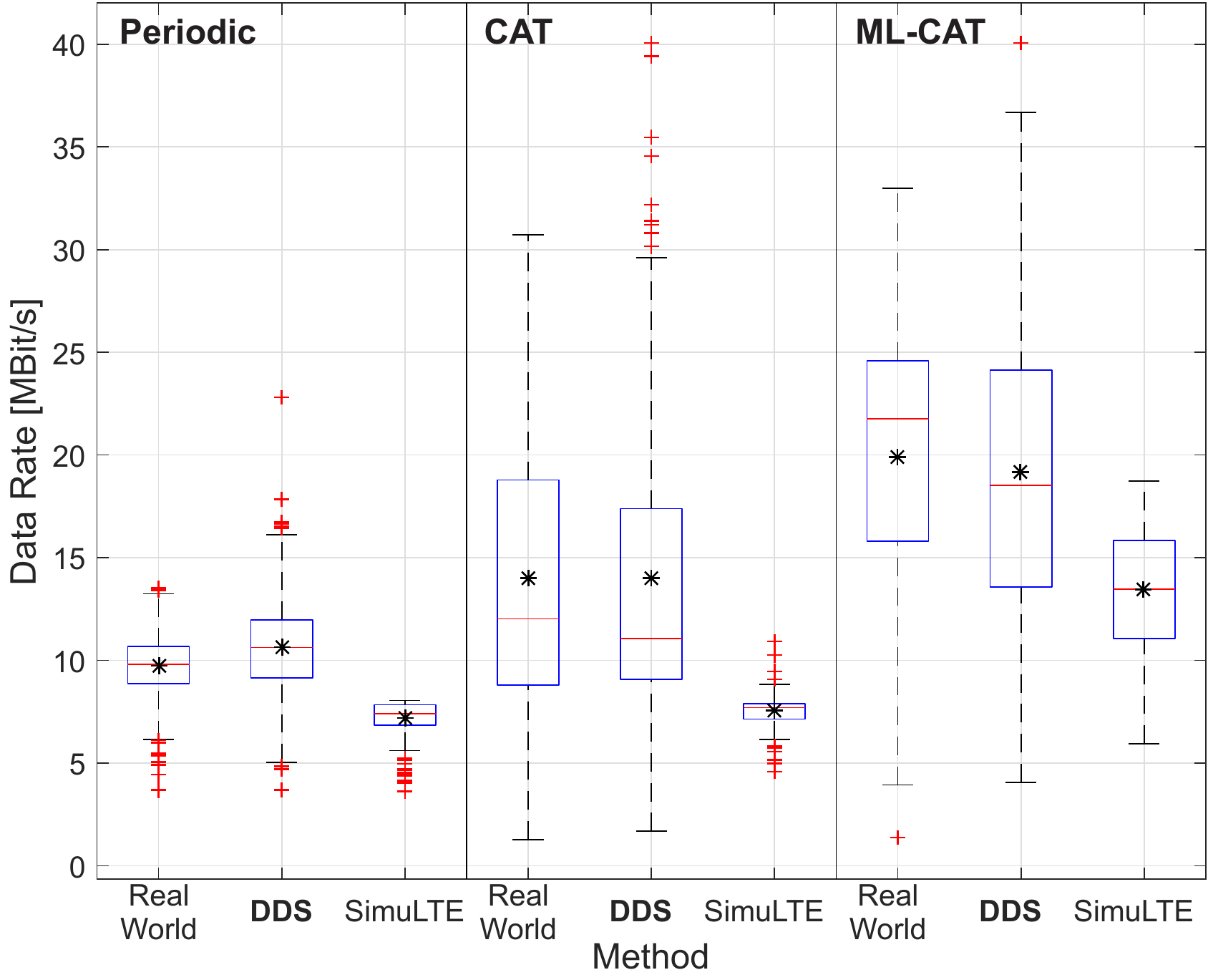}{0.49}{Suburban evaluation track}
	\subfig{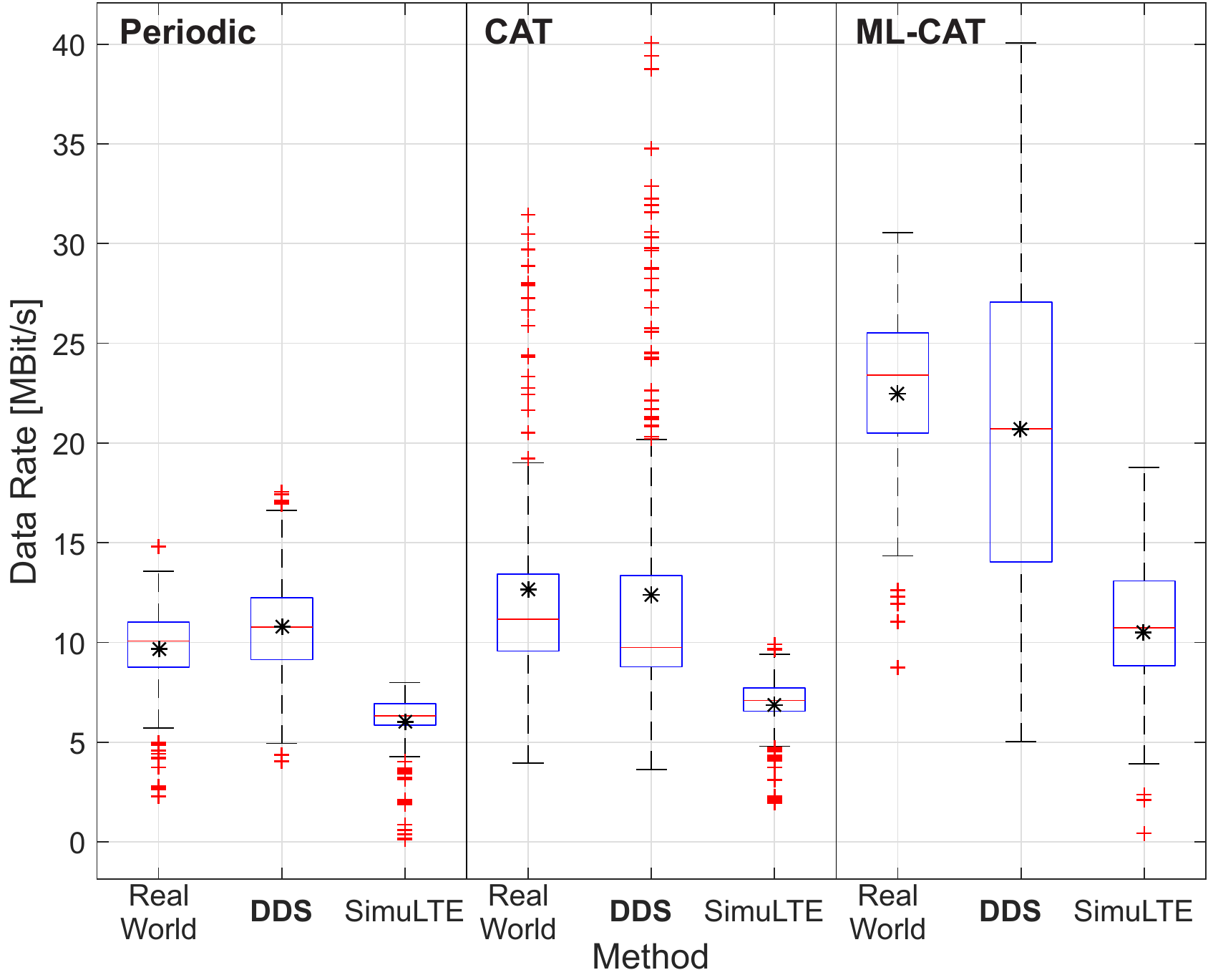}{0.49}{Highway evaluation track}
	
	\caption{Comparison of different performance evaluation methods for the considered \acp{MUS}.}
	\label{fig:cat_validation}
	\vspace{-0.3cm}
\end{figure*}
The resulting end-to-end behavior for the considered methods and transmission schemes is shown in Fig.~\ref{fig:cat_validation}.
%
%
For the real world behavior, several effects can be observed. 
%
%
In contrast to the periodic scheme, \ac{CAT} and \ac{ML-CAT} are able to achieve significant increases in the resulting data rate due to the opportunistic data transmission approach
%
%
The schemes behave differently with respect to the evaluation tracks. While the \ac{SINR}-based \ac{CAT} approach works better on the suburban track, \ac{ML-CAT} achieves the highest benefits on the highway track.

%
%
The proposed \ac{DDS} is able to capture most of these phenomenons very precisely. Especially on the suburban track, the results match well with the real world measurements.
%
%
As discussed in previous work \cite{Sliwa/etal/2018a}, the velocity-related network dynamics lead to a reduced prediction accuracy on the highway track, which results in an increased variance within the \ac{DDS}. This effect has a doubled impact on \ac{ML-CAT}, as the data rate prediction is executed twice (the first time within the transmission scheme itself and the second time within the \ac{DDS}). However, the mean behavior of the data rate is still captured precisely.

%
%
In the SimuLTE evaluation, the effects of the implemented \acp{MUS} are less significant. For the suburban track, \ac{CAT} only increases the mean data rate by $4~\%$ ($44~\%$ in the real world) and \ac{ML-CAT} achieves a boost by $86~\%$ ($105~\%$ in the real world).
During the real word measurements, the observed channel quality is significantly more dynamic than the channel model-based approach of the simulation.
%
%
In addition, SimuLTE does not implement \ac{LTE} \ac{TPC}, which explains the low variance of the results of each scenario.

%
%
It can be argued that a better congruency of simulation results and real world measurements can be achieved by increasing the modeling granularity within the simulation setup. However, in many cases the required parameters are either not easily accessible (e.g., operator internals) or impossible to obtain (e.g., the behavior of all other cell users). Therefore, the required modeling effort is massively increased for those intentions.
In contrast to that, the strength of \ac{DDS} is its ability to extract the parameterization implicitly from the data itself.

\subsubsection{Computation Time}

%
%
In addition to the achievable result validity, the computation time is another crucial factor for the simulative method. Often very detailed simulation environments lead to very high computation times, which limits the applicability for large-scale evaluations.
%
%
\begin{figure}[]  	
	\centering		  
	\includegraphics[width=1\columnwidth]{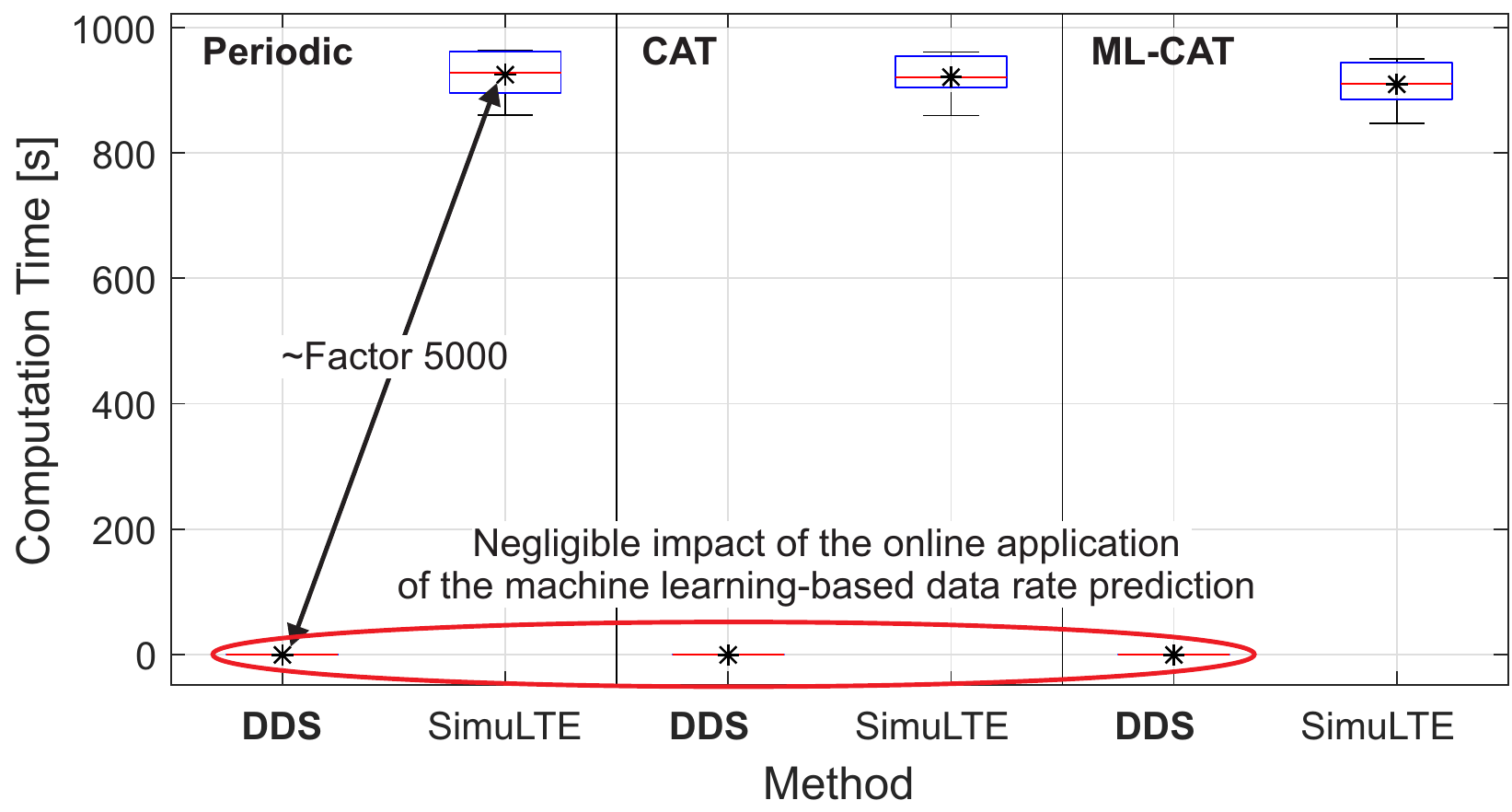}
	\vspace{-0.7cm}	
	\caption{Comparison of different performance evaluation methods for the considered \acp{MUS}.}
	\label{fig:evaluation_speed}
	\vspace{-0.3cm}	
\end{figure}

Fig.~\ref{fig:evaluation_speed} shows the resulting computation times per run for the proposed \ac{DDS} and the system-level network simulator.

In the considered case study scenarios, the proposed \ac{DDS} is more than 5000 times faster than the classical system-level simulation while being significantly more precise. Although the resulting \ac{RF} model is very large -- the generated \text{RF} contains 120533 leafs -- it can be evaluated very resource efficiently, the computation time for online application of the trained model is negligible. 

Within the case study, only the results of a single end-to-end indicator are considered. However, in contrast to the black box approach of the proposed \ac{DDS}, classical network simulation spends a massive effort to derive the corresponding results. In addition to the focused \ac{UE} and its protocol stack, the whole network infrastructure is simulated.

\section{Limitations of \ac{DDS}} \label{sec:limitations}

With this work, \ac{DDS} is spotlighted as sophisticated alternative to classical system analysis approaches for specific evaluation tasks. Although the proposed approach has significant advantages in comparison to other scientific evaluation methods, it has limitations that are related to its data-driven nature.

%
%
Due to the high dependency to the training set, the possible evaluations are fixed to the measured labels. In most cases, introducing additional performance indicators requires to perform additional measurements in the real world. Therefore, the exploitation of expert knowledge about the to be analyzed indicators is of tremendous importance to proactively optimize the data acquisition phase.
%
%
For each considered performance indicator, dedicated prediction and error models need to be learned. In contrast to system-level simulation, the protocol stack is treated as a \emph{black box}, therefore it is not possible to investigate intermediate indicators, which are not contained in the feature set.

\section{Conclusion}

%
%
In this paper, we presented a data-driven approach for end-to-end network performance evaluation of novel communication methods. The proposed scheme -- which is referred to as \ac{DDS} -- has been validated against real world measurements and system-level network simulation. 
%
%
\ac{DDS} exploits the unique characteristics of different machine learning models, which are applied to derive a prediction model and a description for the derivations between predictions and real world measurements. The joint consideration of these models leads to a close to reality evaluation environment for the study of novel communication methods.
%
%
In contrast to system-level network simulation, the scenario modeling overhead is very low as the parameterization is implicitly extracted from the data itself. Moreover, within an example case study, it was shown that \ac{DDS} achieves a significantly better modeling accuracy of the real world behavior while only requiring a fraction of the computation time of regular simulation methods.
%
%
In future work, we will evaluate the applicability of novel machine learning methods for both prediction and simulation model generation. Furthermore, we will utilize \ac{DDS} to analyze novel anticipatory communication methods.
\section*{Acknowledgment}

\footnotesize
Part of the work on this paper has been supported by Deutsche Forschungsgemeinschaft (DFG) within the Collaborative Research Center SFB 876 ``Providing Information by Resource-Constrained Analysis'', project B4.

\bibliographystyle{IEEEtran}
\bibliography{Bibliography}

\end{document}